\documentclass[referee]{raa}            

\usepackage{graphicx,times}             
\usepackage{natbib}
\usepackage{amssymb,amsmath}
\bibpunct{(}{)}{;}{a}{}{,}

\usepackage{CJKutf8}

\usepackage{hyperref}
\hypersetup{hypertex=true,
	colorlinks=true,
	linkcolor=blue,
	anchorcolor=blue,
	citecolor=blue}

\usepackage{graphics}
\usepackage{tipa}
\usepackage{pifont}
\usepackage{subcaption}
\usepackage{mathrsfs}

\usepackage{txfonts}

\usepackage{longtable}
\usepackage{dcolumn}
\usepackage{booktabs}

\usepackage{threeparttable}
\usepackage{xcolor}

\usepackage{booktabs}
\usepackage{multirow}

\usepackage{array}

\newcommand{\resultOmegabForFan}{0.043^{+0.005}_{-0.006}}
\newcommand{\resultOmegabForBosman}{0.045^{+0.004}_{-0.006}}
\newcommand{\resultOmegabForBosmanNonPara}{0.048^{+0.001}_{-0.003}}
\newcommand{\resultOmegabForBosmanFour}{0.048^{+0.004}_{-0.007}}
\newcommand{\resultOmegabForBosmanFive}{0.049^{+0.004}_{-0.006}}

\begin{document}

  \title{Constraints on Baryon Density from the Effective Optical Depth of High-Redshift Quasars

}
   \volnopage{Vol.0 (20xx) No.0, 000--000}      
   \setcounter{page}{1}          

   \author{
   	Wen-Fei Liu 
      \inst{1}
      \thanks{These authors contributed equally to this work.}
   \and Yuan-Bo Xie
      \inst{2,3}
     \footnotemark[1]
   \and Zhi-E Liu
      \inst{1}
    \and Jin Qin
    \inst{3}
    \and {Kang Jiao}
    \inst{3}
    \and {Dong-Yao Zhao}
    \inst{4}
    \and {Tong-Jie Zhang (\begin{CJK}{UTF8}{gbsn}张同杰\end{CJK})}
    \inst{2,3}
   }

   \institute{College of Physics and Electronic Engineering, Qilu Normal University, Jinan, 250200, China\\            
        \and
        Institute for Frontiers in Astronomy and Astrophysics, Beijing Normal University, Beijing 102206, People's Republic of China
        \and
        School of Physics and Astronomy, Beijing Normal University, Beijing 100875, People's Republic of China; {\it tjzhang@bnu.edu.cn}\\
        \and
             Beijing Planetarium, Beijing Academy of Science and Technology, Beijing, 100044, China\\
\vs\no
   {\small Received 20xx month day; accepted 20xx month day}}

\abstract{ 	We present constraints on the baryonic matter density parameter, $\Omega_b$, within the framework of the $\Lambda$CDM model. Our analysis utilizes observational data on the effective optical depth from high-redshift quasars. To parameterize the photoionization rate $\Gamma_{-12}$, we employ a B\'{e}zier polynomial. Additionally, we approximate the Hubble parameter at high redshifts as $H(z)\approx 100h\Omega_m^{1/2} (1+z)^{3/2}$ km s$^{-1}$ Mpc$^{-1}$. Confidence regions are obtained with $h=0.701\pm0.013$ and $\Omega_m = 0.315$, optimized by the \textit{Planck} mission. The best-fit values are $\Omega_b =\resultOmegabForFan$ and $\Omega_b = \resultOmegabForBosman$, corresponding to an old data set and a new data set, respectively. And we test the non-parametric form of $\Gamma_{-12}$, obtaining $\Omega_b = \resultOmegabForBosmanNonPara$. These results are consistent with the findings of \textit{Planck} at the 1 $\sigma$ confidence level. Our findings underscore the effectiveness of quasar datasets in constraining $\Omega_b$, eliminating the need for independent photoionization rate data. This approach provides detailed cosmic information about baryon density and the photoionization history of the intergalactic medium.
}

%
\maketitle
\section{Introduction}
\label{sec:intro}

Quasars are one of the most distant celestial objects that can be observed as luminous sources, often at very high redshifts, with some even approaching the epoch of reionization. This epoch occurred when the first generation of stars and quasars ionized the neutral intergalactic medium (IGM) and ended the cosmic "dark ages" \citep{paper:16}. The growing number of observations of high-redshift quasars has provided valuable measurements and made them an effective probe for studying the early universe. Keck and VLT spectroscopy of high-redshift quasars \citep{paper:02,paper:15} have shown the first observation of a complete Gunn-Peterson trough in the spectrum of the $z=6.28$ quasar SDSS 1030+0524. \cite{2015_Becker_extreme_Lya_trough}  ave discovered an exceptionally long and dark Ly$\alpha$ trough that extends to high redshifts. This trough represents the longest one found to date below a redshift of 6. \cite{paper:04} observed a particularly dark region of length approximately 5 Mpc at $z\approx5.4$ along the line of sight to the $z\approx5.8$ quasar SDSS 1044-0125. Observations of high-redshift quasars by \cite{paper:05}, \cite{paper:02}, and \cite{paper:15} have shown that the Ly$\alpha$ absorption due to neutral hydrogen in the IGM increases dramatically toward high redshifts.

One of the primary goals of cosmology is to constrain cosmological parameters using various observational quantities that depend on redshift. In this paper, we utilize measurements of the effective optical depth obtained from spectroscopic observations of high-redshift quasars, spanning a redshift range of $z=4.8$ to $6.2$, to constrain the baryon density parameter, $\Omega_b$. Some of these data sets have previously been employed to investigate the evolution of the ionizing background and the epoch of reionization \citep{paper:06,paper:07}. However, this study represents the first application of these data sets to constrain $\Omega_b$, which is a crucial cosmological parameter commonly determined through the Cosmic Microwave Background (CMB) or the Baryon Acoustic Oscillation (BAO) measurements from galaxy clustering. While these traditional methods are applicable at very high redshifts and relatively low redshift epochs, respectively, our work provides an important complement by utilizing high-redshift quasars to derive constraints on $\Omega_b$ within the redshift range between the CMB and galaxies.

In Section \ref{sec:ob}, we present the observational data. The derivations of the effective optical depth and the constraint methods are introduced in Section \ref{sec:meth}. Finally, we will discuss and conclude in Section \ref{sec:discussion}.

\section{Observational data}
\label{sec:ob}

In this work, we selected the data of effective optical depth for Ly$\alpha$ from \cite{paper:07} and the up to date sample from \cite{2022_Bosman_effective_optical_depth} to process cosmological constraints. 

For all the specra in the work of \cite{paper:07}, twelve of the spectra were obtained using the Keck ESI instrument, while the remaining spectra were observed using the MMT Red Channel and Kitt Peak 4 m MARS spectrographs. The data used in this study have a spectral resolution of approximately $R \sim 3000-6000$, which depends on the seeing conditions. They binned all the Keck ESI spectra to a resolution of $R = 2600$. Although the signal-to-noise ratios (S/Ns) of these spectra vary by a factor of approximately 7, they have obtained sufficiently long observations for all quasars at redshifts greater than 6.1 to ensure accurate measurement of complete Gunn-Peterson troughs. The nonuniform S/N among the lower redshift quasars does not significantly affect the analysis. In fact, uncertainties in the average transmission are primarily influenced by large sample variance\citep{paper:07}.

\begin{figure}
	\centering
	\includegraphics[width=0.7\textwidth]{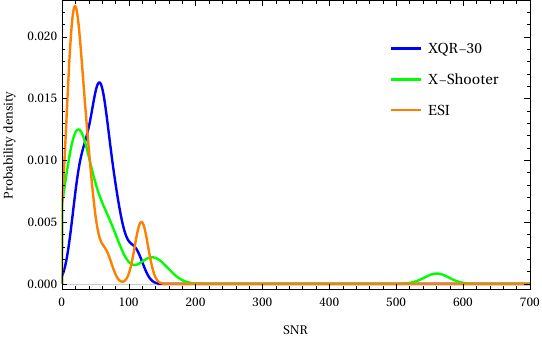}
	\caption{The probability distribution functions of SNR for three different datasets used to construct \cite{2022_Bosman_effective_optical_depth} quasar data: XQR-30 (blue), X-Shooter (green), and ESI (orange). It is observed that the SNR distribution for most of the data lies between 10 and 100.  }
	\label{2022_Bosman_SNR}
\end{figure}

The dataset used in the study by \cite{2022_Bosman_effective_optical_depth} includes 30 quasar spectra at $ z\gtrsim 5.8$ from the XQR-30 program (1103.A-0817(A)), 26 archival spectra obtained with the X-Shooter spectra of equal SNR $>$10 per 10 ${\rm km\ s^{-1}}$ pixel, and 16 archival spectra of quasars with $ \gtrsim 5.7$ obtained with the ESI instrument on the Keck Telescope. All XQR-30 spectra have signal-to-noise ratios (SNRs) greater than 20 per 10  ${\rm km\ s^{-1}}$ pixel within the wavelength range of 1165 $\AA$ to 1170 $\AA$. The X-Shooter instrument provides a resolution of approximately 34 $\rm{km\ s^{-1}}$ in the visible range (5500 $\AA$ to 10200 $\AA$) and approximately 37 $\rm{km\ s^{-1}}$ in the infrared range (10200 $\AA$ to 24800 $\AA$). However, due to better-than-average seeing conditions during observations, the effective resolution is slightly higher. On the other hand, the ESI instrument has a lower spectral resolution of approximately 60 $\rm{km\ s^{-1}}$ and its wavelength coverage is limited to the optical range up to $\lambda<10500$ $\AA$. Figure \ref{2022_Bosman_SNR} illustrates the distribution of signal-to-noise ratios (SNRs) for quasars obtained from the XQR-30, X-Shooter, and ESI datasets. The figure clearly demonstrates that, in general, the XQR-30 dataset exhibits higher SNR values. All the quasar samples were divided into 14 bins, and the mean Ly$\alpha$ transmission was obtained for each bin. 

In this study, we utilize the data from both \cite{paper:07} and the dataset provided by \cite{2022_Bosman_effective_optical_depth} to constrain the value of $\Omega_b$. Our analysis yields similar results that are in agreement with the findings of \cite{2018_planck_results} within the 1 $\sigma$ confidence region.

\section{Methodology}
\label{sec:meth}

The effective optical depth are derived with the consideration that the intergalactic medium (IGM) is inhomogeneous. This assumption is reasonable because the actual distribution of the IGM in the universe is complex and inhomogeneous, and therefore would necessarily affect the observations. Additionally, another consideration in the derivation of the theoretical effective optical depth is the approximation of the Hubble parameter at high redshifts. When $z\gg1$, the Hubble parameter $H(z)$ can be approximated as $H(z)\approx 100h\Omega_m^{1/2} (1+z)^{3/2}$ km ${\rm s}^{-1}$ ${\rm Mpc}^{-1}$. This approximation is advantageous because the observational data are all from quasars at very high redshifts ( $z=4.8\thicksim6.2$), some of which even approach the reionization epoch. We will provide detailed derivations for the effective optical depth below.

\subsection{Effective optical depth for Ly$\alpha$}

Assuming an approximate thermal equilibrium between photoionization heating by the UV background and adiabatic cooling due to Hubble expansion \citep{paper:09,paper:11}, the optical depth for Lyman-$\alpha$ and Lyman-$\beta$ absorption lines, also known as the Gunn-Peterson optical depth, depends on the local density of the IGM. Therefore, the non-uniformity of the IGM must be taken into account. The fractional density of the IGM is defined as $\Delta \equiv \rho/\langle \rho \rangle$ $\equiv \delta + 1$ (where $\delta$, the density contrast, is the departure of the local density from the mean density, in units of the mean density). For a region of IGM with density $\Delta$, the effective optical depth can be written as~\citep{paper:06}:
\begin{equation}
	\tau(\Delta)\propto \frac {(1+z)^6(\Omega_bh^2)^2 \alpha(T)} {\Gamma(z)
		H(z)} \Delta^2 ,
	\label{1}
\end{equation}
where $\Omega_b$ is the baryon density, $\Gamma(z)$ is the photoionization rate, and $\alpha(T)$ is the
recombination coefficient at temperature $T$~\citep{paper:01},
$\alpha(T) = 4.2\times10^{-13} (T/10^4\rm K)^{-0.7} {\rm cm}^{3} {\rm s}^{-1} .$

The dependence on $\Delta^2$ comes from $\tau \propto n_{\rm
	HI}$~\citep{paper:08}, which is proportional to $n_{\rm H}^2$~\citep{paper:09}, and
proportional to $\Delta^2$ for a highly ionized IGM~\citep{paper:06}.
The temperature of the IGM is determined by
photoionization-recombination equilibrium, which leads to a
power-law relation between temperature and density in the form of $T =
	T_0 \Delta^{\gamma-1}$, with $T_0\sim 1-2\times10^4$ K, and $\gamma$ ranges from -1 to 0 \citep{paper:11}.

The optical depth for Ly$\alpha$ can be expressed as

\begin{equation}
	\tau_\alpha(\Delta)=\tau_{0} \frac{(1+z)^6}{\Gamma_{-12}(z)}
	\frac{(\Omega_bh^2)^2}{H(z)} \Delta^2 \left(\frac{2}{T(z)}\right)^{0.7},
	\label{2}
\end{equation}

where $\Gamma_{-12}(z)$ is the photoionization rate in units of $10^{-12}$ s$^{-1}$, whose data matching
the reshifts of samplers can be got from Fan et al.~\citep{paper:06,paper:07}. The numerical constant $\tau_0$ is determined below. 
$T(z)$ is temperature in units of $10^4$K, which will be discussed in section 3.3.

At high redshift $H(z)\approx 100h\Omega_m^{1/2} (1+z)^{3/2}$ km s$^{-1}$ Mpc$^{-1}$, and
the equation (2) can be
rewritten as

\begin{equation}
	\tau_\alpha(\Delta)=\tau_1 \frac{(1+z)^{4.5}}{\Gamma_{-12}(z)}
	\frac{\Omega_b^2}{\Omega_m^{0.5}} h^3 \Delta^2 \left(\frac{2}{T(z)}\right)^{0.7} ,
	\label{3}
\end{equation}

where numerical constant $\tau_1=\tau_0 /(3.24\times10^{-18})$ (in CGS unit
system). We follow McDonald \& Miralda-Escud\'{e}
~\citep{paper:13} and set $\tau_1=0.664$, {corresponding to $\tau_0 = 2.15\times 10^{-18} $}.

Eq.~\eqref{3} is appropriate for all Lyman series lines, with
different values of the proportionality constant~\citep{paper:07}.

The observed transmitted flux ratio $\mathscr{T}$ is averaged
over the entire IGM density distribution~\citep{paper:06}
\begin{equation}
	\mathscr{T} = \langle e^{-\tau} \rangle = \int_0^\infty
	e^{-\tau(\Delta)} p(\Delta) d(\Delta),
	\label{4}
\end{equation}

This equation assumes that the universe is fully ionized. However, several recent studies \cite{2019_Kulkarni_not_fully_ionized_>5.5} have pointed out that this assumption may not hold true at $z>5.5$. If the findings of \cite{2019_Kulkarni_not_fully_ionized_>5.5} are accurate, the optical depth will be larger. In order to balance equation \eqref{3}, a larger value of $\Omega_b$ is required. 
While we recognize that the assumption of full ionization may not be entirely valid in the redshift range $z=4.8\text{ to } 6.2$, the intergalactic medium (IGM) is predominantly ionized in this range, meaning that the assumption of full ionization remains approximately valid. Furthermore, the impact on the constraints of $\Omega_b$ falls within the acceptable error margins of the parameter limits. To simplify the analysis and maintain consistency with previous studies, we continue to assume a fully ionized universe. It is important to note that our analysis is based on observational data, which generally supports the assumption of full ionization in this redshift range, as demonstrated by \cite{paper:06,paper:07}. However, future research may benefit from incorporating models with partial ionization, particularly at higher redshifts.

The distribution function of the intergalactic medium's (IGM) density, denoted as $p(\Delta)$, is a function of the volume-weighted density distribution \citep{paper:12}.

\begin{equation}
	p(\Delta) = A \exp \left[ - \frac{(\Delta^{-2/3} - C_0)^2}
	{2(2\delta_0/3)^2}\right] \Delta^{-\beta},
	\label{5}
\end{equation}

The parameter $\delta_0 = 7.61/(1+z)$, where $z$ is the redshift, and the constants $\beta$, $C_0$, and $A$ are numerical values listed in Table 1 of Miralda-Escud'e et al. \citep{paper:12} for various redshifts. These constants are obtained from numerical simulations, which align well with the observational results of the Lya forest transmitted flux \citep{paper:12,1997_Rauch}.

The definition of effective optical depth is
\begin{equation}
	\tau_{\rm GP}^{\rm eff}  \equiv - \ln({\mathscr{T}}),\label{6}
\end{equation}
Combining the Eq.~\eqref{3}, \eqref{4}, \eqref{5} and~\eqref{6}, one can get the
final equation for calculating the Ly$\alpha$ theoretical effective optical depth
for inhomogeneous IGM with high redshift approximation of Hubble
parameter.

\begin{figure}
	\centering
	\includegraphics[width=0.7\textwidth]{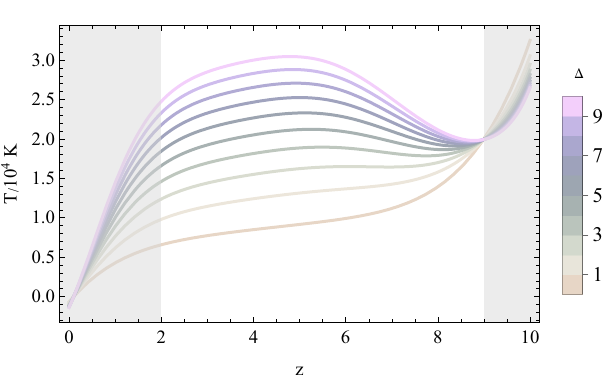}
	\caption{The thermal history, $T$, with respect to different density $\Delta$, ranges from 1 to 10. The gray areas represent the regions where $z < 2$ and $z > 9$, with no $T_0$ provided in \cite{2015_Matthew_thermal_history_Figure2}; therefore, $T$ may not be reliable in those regions. However, the central region has already covered the redshift range of the data used in this paper. We can observe that when the density is low, the temperature increases monotonically with redshift. However, when the density is high, the temperature approximately reaches a maximum at $z = 5$, and then decreases.}
	\label{thermal history}
\end{figure}

\subsection{Parameterisation of the photoionization rate $\Gamma_{-12}$}

In this work, we employ a B\'{e}zier fit to describe the evolution of $\Gamma_{-12}$ with redshift. The parameterization is given by 
\begin{equation}
	\Gamma_{-12}(z) =\sum_{d=0}^n a_d \frac{n !\left(z / z_m\right)^d}{d !(n-d) !}\left(1-\frac{z}{z_{\mathrm{m}}}\right)^{n-d}
	\label{8}
\end{equation}

where $z_m$ is the maximum redshift in the quasar data, and $a_d$ are the coefficients. This method, first introduced by Amati et al.~\citep{2019_Amati}, has been previously used to reconstruct Hubble data. We set $n=2$ to balance the trade-off between the flexibility of the parameterization curve, determined by the number of coefficients $a_d$, and the precision of the fitting results, which is limited by the number of available data points. This parameterization take the advantage of the flexibility of B\'{e}zier curves compared with other curves like polynomial. 
The flexibility of B\'{e}zier is more advantageous for parametric constraints than other curves, such as polynomials.

\subsection{Thermal history parameters}

We refer to Figure 2 in \cite{2015_Matthew_thermal_history_Figure2} to establish a reasonable form for $T(z)=T_0(z)\Delta^{\gamma(z)-1}$ . The curve depicted in Figure 2 of \cite{2015_Matthew_thermal_history_Figure2} can be well approximated by the following two quintic polynomials.

\begin{equation}
	\begin{aligned}
		{T_0(z)=0.000276332 z^5-0.0059786 z^4+0.0541698 z^3-0.257399 z^2+0.703909 z-0.0692726}\\
		{\gamma(z)=0.000190549 z^5-0.00480521 z^4+0.0436158 z^3-0.187106 z^2+0.369498 z+1.308008}
	\end{aligned}
\end{equation}

Figure \ref{thermal history} displays the thermal history for different densities ranging from 1 to 10. It is evident that within our region of interest $z\sim(4.8-6.2)$, the form of $T(z)$ will have a significant impact on the optical depth $\tau_\alpha(\Delta)$.

\subsection{Fitting methods}

Considering that the data are independently and identically distributed as Gaussian, we have a likelihood form as follows

\begin{equation}
	\mathscr{L} = \prod_{i=1}^{n} \frac{1}{\sqrt{2\pi\sigma_i^2(z_i)}} {\rm exp}\left[-\frac{\left(\tau^{eff}_{th}(
		\Omega_b, a_0,a_1,a_2,\sigma_{int}\ |\ z_i)-\tau_{ob}^{eff}(z_i)\right)^2}{2\sigma_i^2(z_i)}\right]
\end{equation}

$\tau^{eff}_{th}$ is the predicted value of effective optical depth,
$\tau_{ob}^{eff}$ is the corresponding observational data. In this paper, we set $\Omega_m = 0.315$,   and $h=0.701\pm0.101$\citep{2018_planck_results}. $a_0, a_1, a_2$ are three parameters related to $\Gamma_{-12}$. Where $\sigma_i(z_i)^2=	\sigma^2_{\tau}(z_i)+\sigma^2_{int}$, $\sigma^2_\tau(z_i)$ is the error in effective optical depth data. $\sigma_{int}$ is the global intrinsic dispersion.

To calculate the posterior distribution of the model parameters, we employed the Affine Invariant Markov Chain Monte Carlo (MCMC) ensemble sampler, emcee \citep{emcee_2013}, implemented in Python. This method allowed us to survey the posterior distribution in the parameter space and maximize the likelihood function. The resulting contour plots were generated using the corner module \citep{corner}.

\begin{figure}
	\centering
	\includegraphics[width=0.85\textwidth]{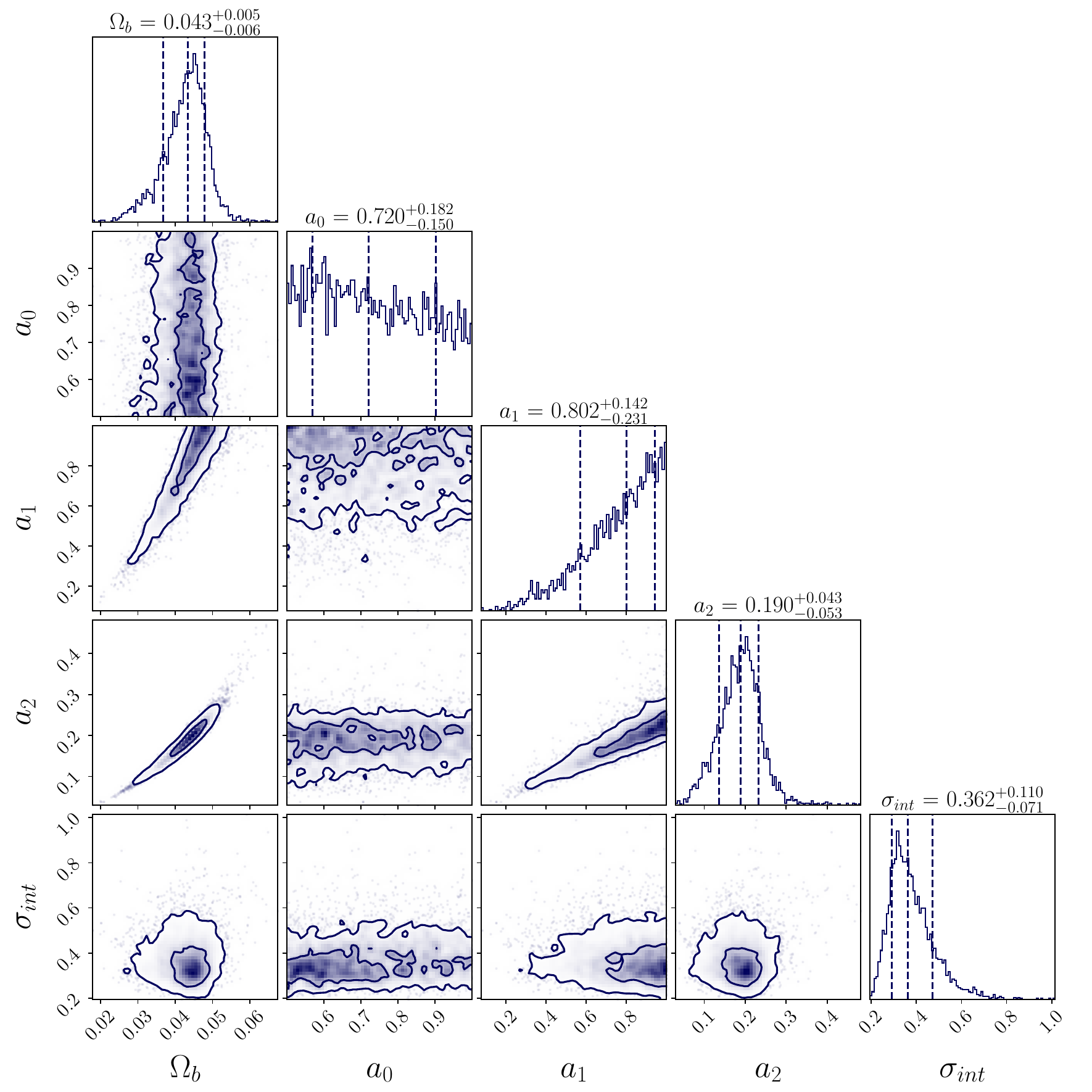}
	\caption{ The 1D and 2D contours representing the 1$\sigma$ and 2$\sigma$ confidence regions for \cite{paper:07} Ly$\alpha$ data. It can be seen that although $a_0$ and $a_1$ cannot be well constrained, $a_2$, especially the most important parameter, the baryon density, has been well constrained. The value of the baryon density is consistent with the results of Planck within a 1 $\sigma$ confidence interval.}
	\label{corner_Fan}
\end{figure}
\section{Discussion and conclusion}
\label{sec:discussion}

\begin{figure}
	\centering
	\includegraphics[width=0.85\textwidth]{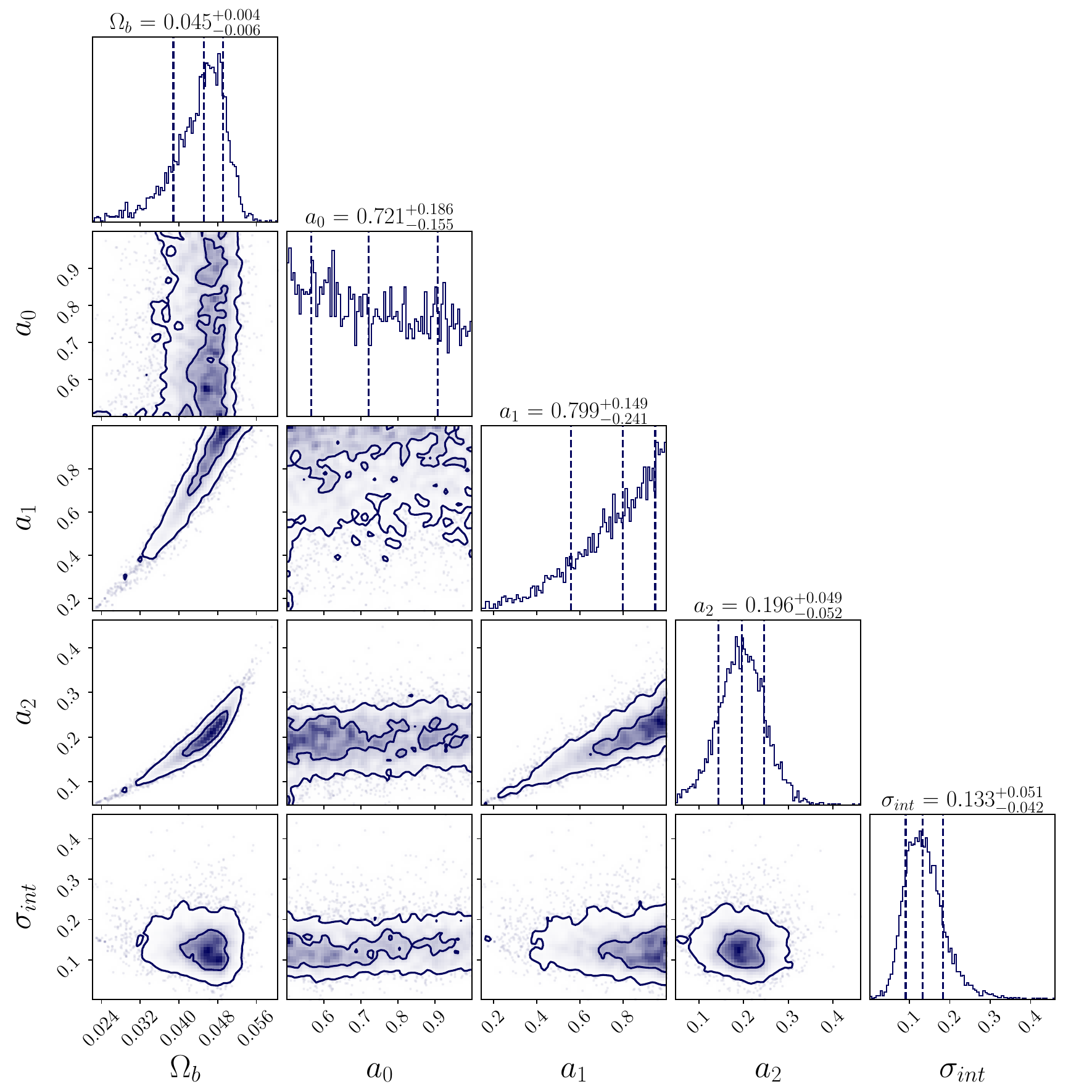}
	\caption{ The 1D and 2D contours representing the 1$\sigma$ and 2$\sigma$ confidence regions for \cite{2022_Bosman_effective_optical_depth} Ly$\alpha$ data. Similar to the results of Figure \ref{corner_Fan}, $a_0$ and $a_1$ are not well constrained, but $a_2$ and the baryon density, $\Omega_b$, are better constrained, and the value of $\Omega_b$ is consistent with the Planck results.}
	\label{corner_Bosman}
\end{figure}

%
%

%
%
%

\begin{figure}[htbp]
	\centering
	\begin{subfigure}[b]{0.48\textwidth}
		\centering
		\includegraphics[width=\textwidth]{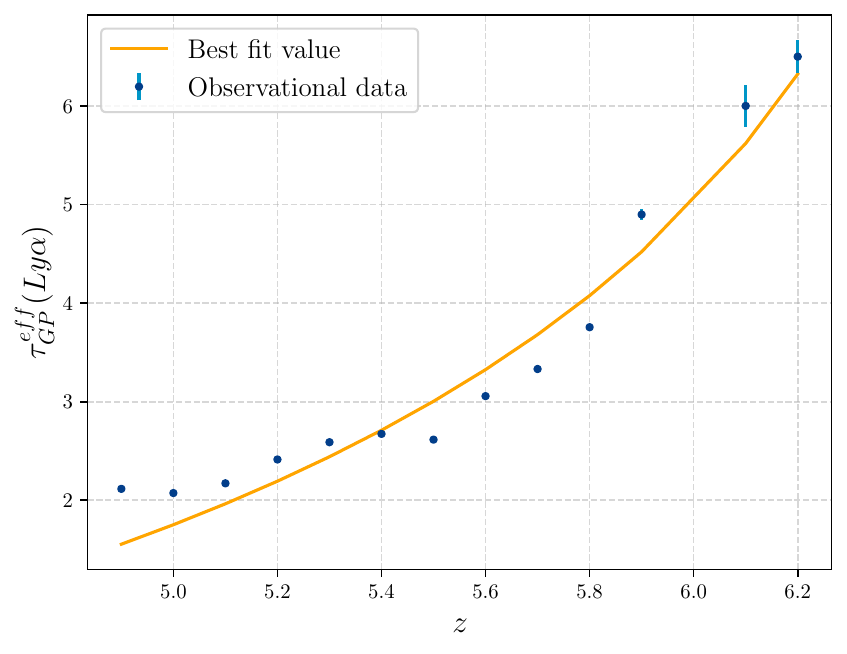}
		\caption{Data from \cite{paper:07}}
	\end{subfigure}
	\begin{subfigure}[b]{0.48\textwidth}
		\centering
		\includegraphics[width=\textwidth]{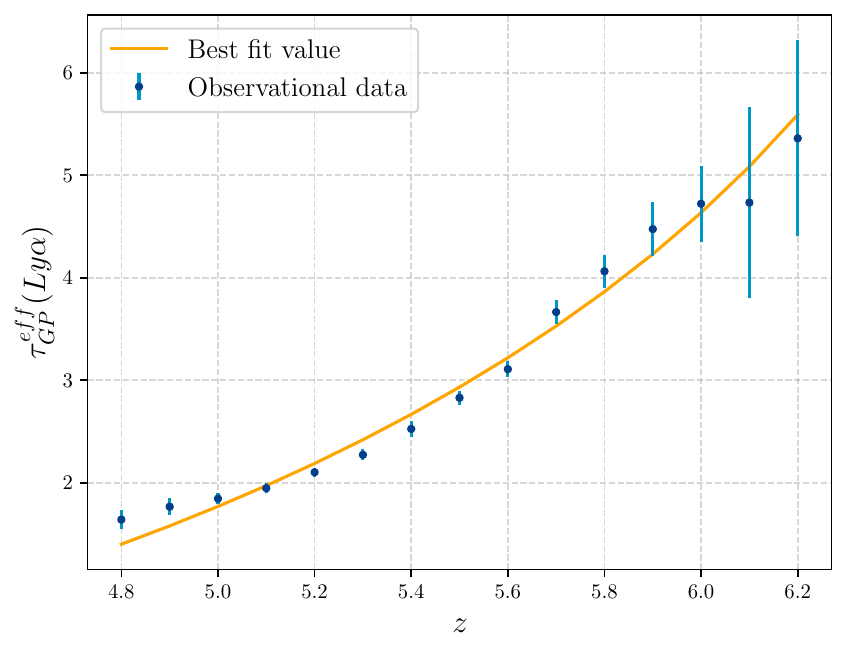}
		\caption{Data from \cite{2022_Bosman_effective_optical_depth}}
	\end{subfigure}
	\caption{Theoretical values and observed effective optical depths for Ly$\alpha$ data. 
		The theoretical values are represented by yellow triangles, while the observed values are shown as blue points with error bars. We can see that the results from the two datasets fit well and are consistent.}
	\label{best_fit}
\end{figure}

\begin{figure}
	\centering
	\includegraphics[width=0.8\textwidth]{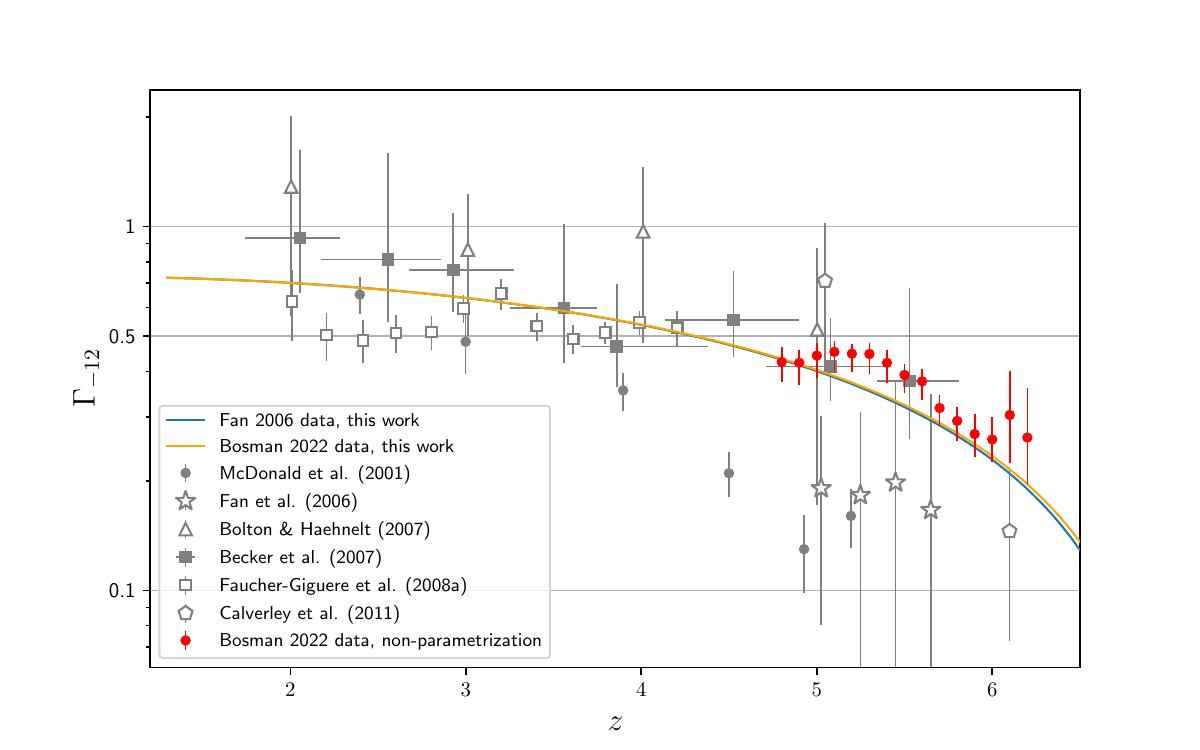}
	\caption{The photoionization rate has been reconstructed along with the 1 $\sigma$ confidence interval. The orange curve represents the results from \cite{paper:07}, while the blue curve corresponds to the findings of \cite{2022_Bosman_effective_optical_depth}. Data with error bars represent previous measurements of $\Gamma_{-12}$. Filled circles denote empirical measurements from the Ly$\alpha$ forest effective opacity by \cite{paper:13}. Stars are from \cite{paper:07}. Triangles are from \cite{2007_Bolton_Gamma-12}. Filled squares are from \cite{2007_Becker_Gamma-12}. Squares are from \cite{2008_Faucher_Gamma-12}. 
	Pentagons represent measurements using the quasar proximity effect by \cite{2011_Calverley_Gamma-12}. Note that all those previous measurements require an assumption on cosmological parameters, such as $h$, $\Omega_m$, and $\Omega_b$. Although such a comparison may not be particularly meaningful, intuitively, our results are consistent with previous measurements that depended on cosmological parameters. And the red dot with the error bar represents the non-parametric form of $\Gamma_{-12}$, which is also consistent with previous work.}
	\label{ionization_rate}
\end{figure}
\begin{figure}[h]
	\centering
	\includegraphics[width=0.95\textwidth]{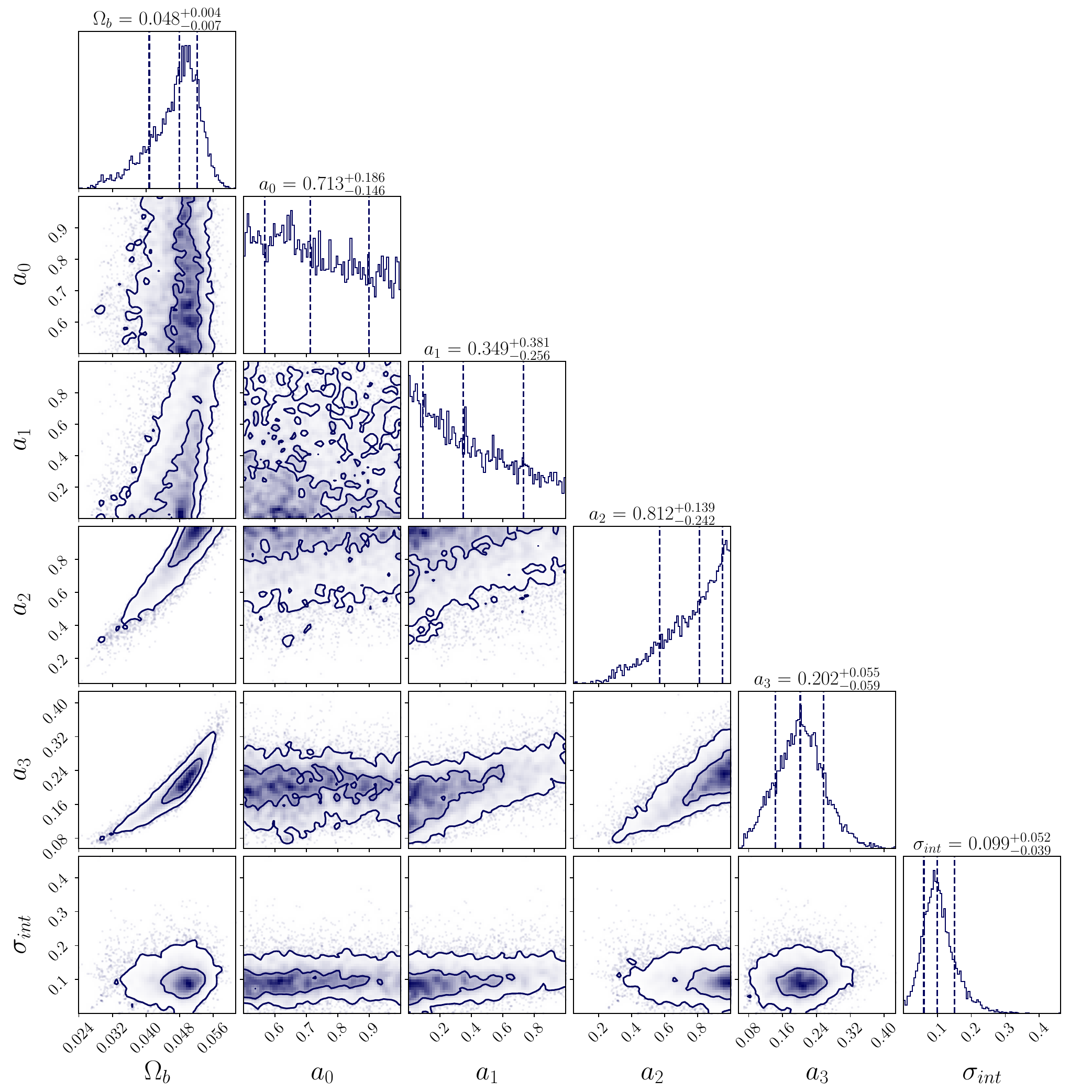}
	\caption{   The 1D and 2D contours represent the 1$\sigma$ and 2$\sigma$ confidence regions for the Ly$\alpha$ data from \cite{2022_Bosman_effective_optical_depth}, using a B\'{e}zier polynomial with 4 parameters. It can be seen that the baryon density obtained with four parameters is slightly higher than that with three parameters, but it is consistent with the three-parameter results within the 1-$\sigma$ confidence interval. }
	\label{test_B_4}
\end{figure}

\begin{figure}[h]
	\centering
	\includegraphics[width=0.95\textwidth]{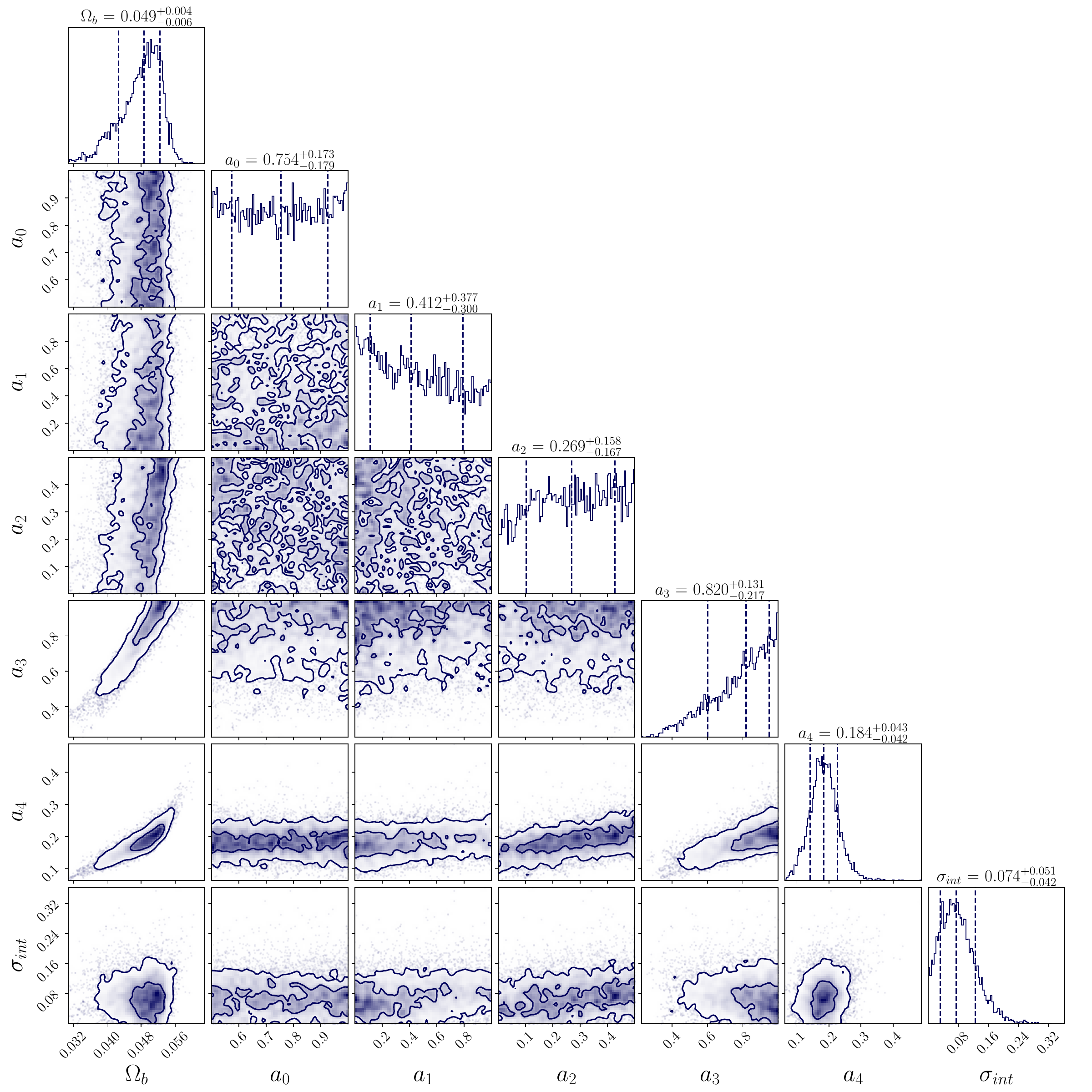}
	\caption{  The 1D and 2D contours represent the 1$\sigma$ and 2$\sigma$ confidence regions for the Ly$\alpha$ data from \cite{2022_Bosman_effective_optical_depth}, using a B\'{e}zier polynomial with 5 parameters. Similar to the four-parameter case, the constraint results for the function parameters are consistent with the three-parameter results within the 1$\sigma$ confidence interval.}
	\label{test_B_5}
\end{figure}

\begin{figure}[h]
	\color{blue}
	\centering
	\includegraphics[width=1\textwidth]{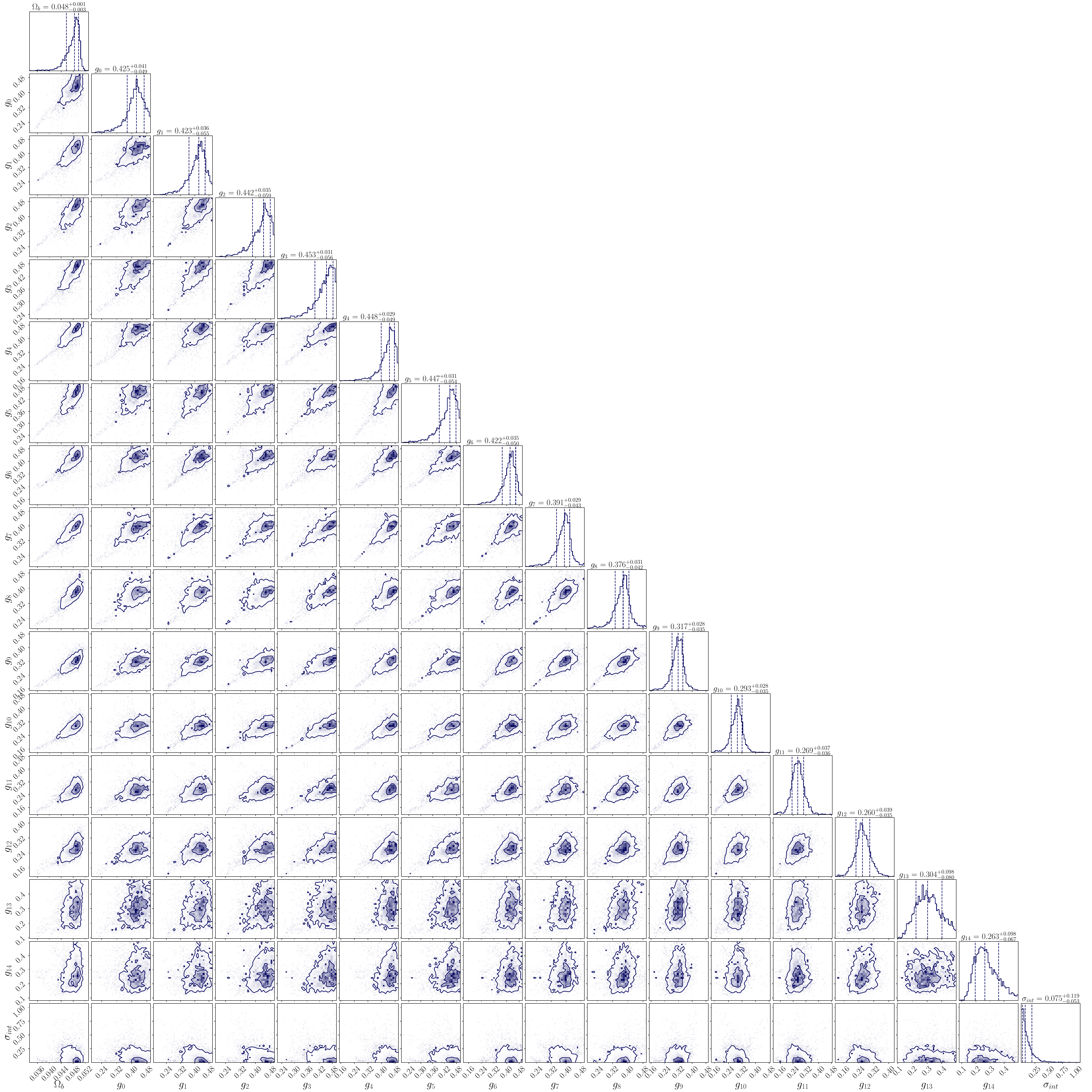}
	\caption{ The 1D and 2D contours represent the 1$\sigma$ and 2$\sigma$ confidence regions for the Ly$\alpha$ data from \cite{2022_Bosman_effective_optical_depth}, using 15 $\Gamma_{-12}$ parameters at each redshift. It is clear that the constraining power remains strong, allowing $\Omega_b$ to be determined with small uncertainties}
	\label{non_para}
\end{figure}

For the first dataset from \cite{paper:07}, the constraining results are shown in Figure \ref{corner_Fan}. Since the data is concentrated in the high redshift range, the parameters $a_0$ and $a_1$ corresponding to the low redshift part of the curve cannot provide a strong constraint. Nevertheless, we have successfully constrained the baryon density, $\Omega_b$, with a best-fit value of $\resultOmegabForFan$. This finding supports a universe that is in agreement with the results from \cite{2018_planck_results}. The 1$\sigma$ confidence region is narrow, indicating high precision in our test results. Additionally, the left panel of Figure \ref{best_fit} shows the Ly$\alpha$ data and its theoretical values with the constrained parameters, while Figure \ref{ionization_rate} displays the reconstructed shape of $\Gamma_{-12}(z)$. We can observe that the photoionization rate decreases with redshift, indicating the presence of abundant neutral hydrogen during the early stages of the universe.

{	
		For the second dataset from \cite{2022_Bosman_effective_optical_depth}, the constraining results are shown in Figure \ref{corner_Bosman}. We obtain a slightly larger value of $\Omega_b=\resultOmegabForBosman$ compared to that of \cite{paper:07}. Those results for $\Omega_b$ are in good agreement with \cite{2018_planck_results} within the 1$\sigma$ confidence region. The right panel of Figure \ref{best_fit} shows the fitting results. 
		
		Figure \ref{ionization_rate} shows the reconstructed ionization rate. It is clear that the results from the two datasets are similar at high redshift but exhibit a larger dispersion at low redshift. This is expected as we are using high redshift data. We also tested the robustness of the parameterization of $\Gamma_{-12}$ using a 4-parameter B\'{e}zier polynomial and a 5-parameter B\'{e}zier polynomial. The test results are shown in Figures \ref{test_B_4} and \ref{test_B_5}, respectively. Compared to the 3-parameter case, the 4- and 5-parameter B\'{e}zier polynomials yield larger values of $\Omega_b = \resultOmegabForBosmanFour$ and $\Omega_b = \resultOmegabForBosmanFive$, respectively, which, however, remain consistent with the Planck results within 1$\sigma$.
		
		A higher baryon fraction would lead to an increased number of electrons available for scattering photons, thereby augmenting the electron scattering optical depth. Furthermore, a greater baryon fraction implies a larger pool of baryons ready for star and galaxy formation. This could potentially stimulate galaxy formation and evolution at higher redshifts. The elevated baryon fraction can also influence feedback processes within galaxies. Feedback mechanisms, such as supernovae explosions and active galactic nuclei, control the growth of galaxies by infusing energy and momentum into the surrounding gas. With a higher baryon fraction, these feedback processes could become more efficient, potentially impacting the gas dynamics, star formation rates, and overall galaxy properties at higher redshifts. This could provide a more direct explanation for the recent observations made by the James Webb Space Telescope (JWST) of more massive galaxies at these redshifts than previously anticipated \citep{2023_Labbe_JWST_more_massive_galaxies_at_high_redshifts}.
		Another possible reason for the larger $\Omega_b$ value might be the relatively limited amount of data. However, this result is still include Planck's results within the 1 $\sigma$ region. Therefore, we must admit that the slight increase in $\Omega_b$ is not sufficient to explain the abundance of bright galaxies at an early age. And the parameterization form has a mild effect on the final results. 
		
	}
	
	{

		We believe that the constraining power on $\Omega_b$ originates from redshift dependence. Equation \eqref{2} demonstrates the degeneracy between $\Gamma_{-12}$ and $\Omega_b$ at a specific redshift. However, by combining data from different redshifts, we could break this degeneracy and place a strong constraint on $\Omega_b$. In such a case, $\Omega_b$ would affect the amplitude of the $\tau_{eff}$ curve, while $\Gamma_{-12}$ would influence both the amplitude and shape of the curve. Therefore, we test the non-parametric form of $\Gamma_{-12}(z)$. We use 15 parameters $g_i$ for $i$ ranging from 0 to 14, each representing $\Gamma_{-12}$ at a specific redshift in the data. Because the data lie in the high-redshift range, we assume these parameters have a flat prior of $(0 \sim 0.5)$. The test results are shown in Figure \ref*{ionization_rate}, and the corner plot for $g_i$ is displayed in Figure \ref{non_para}. From the plots, we can see that the baryon density parameter is well constrained: $\Omega_b = \resultOmegabForBosmanNonPara$. The constraining power remains strong even with the non-parametric form of $\Gamma_{-12}$.
	}

\begin{figure}[h]
		\centering
		\includegraphics[width=1\textwidth]{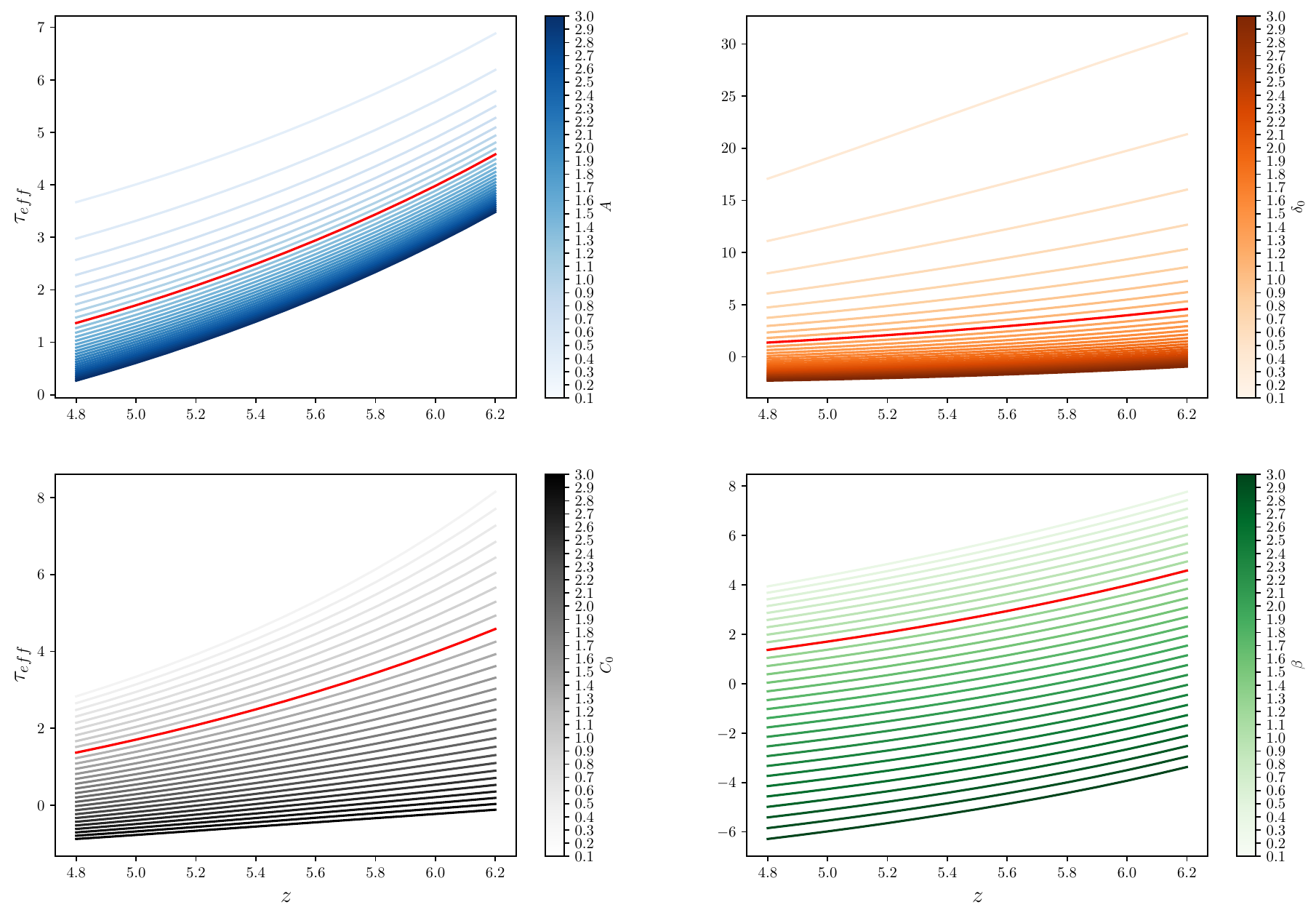}
		\caption{  The influence of density PDF parameters $A$, $\delta_0$, $C_0$, and $\beta$ on $\tau_{eff}$ is substantial. Each parameter was adjusted by a factor ranging from 0.1 to 3. It is evident that those parameters will significantly affect $\tau_{eff}$.Therefore, our result depends on the robustness of those parameters.}
		\label{test_PDF}
	\end{figure}

	Another important factor that could dramatically influence the results is the distribution function of the intergalactic medium's density, which is shaped by equation \eqref{5}. We examined the influence of the four parameters on $\tau_{eff}$ and displayed the results in Figure \ref{test_PDF}. It is evident that these variations in the four parameters significantly affect $\tau_{eff}$. Therefore, the robustness of our results partially depends on the accurate and comprehensive understanding of the IGM properties, especially the density distribution of the IGM.

	In conclusion, this study represents the first effort to use two sets of effective optical depth data, obtained from high-redshift quasars, to constrain the cosmological parameter $\Omega_b$. We found that incorporating photoionization rate data is not crucial, as the parameterization of $\Gamma_{-12}$ proves effective when ample data is available. Despite the limited dataset, our analysis yielded highly precise results with minimal uncertainties, showing good consistency with \cite{2018_planck_results}. Our work is valid for high-redshift $z\gtrsim 4.5$. A high precision of thermal history and the correct shape of the density distribution are crucial for constraining $\Omega_b$. Future research will aim to refine the constraints on $\Omega_b$ by combining these findings with data from low-redshift celestial objects such as supernovae, the Hubble parameter, and baryon acoustic oscillations (BAO). This integrated approach promises to further enhance our understanding of $\Omega_b$.

	%
%
%
%

\normalem
\begin{acknowledgements}
We are grateful for the insightful and useful comments of the referee that greatly helped us improve our manuscript. T.J.Z. (\begin{CJK}{UTF8}{gbsn}张同杰\end{CJK}) dedicates this paper to the memory of his mother, Yu-Zhen Han (\begin{CJK}{UTF8}{gbsn}韩玉珍\end{CJK}), who passed away 4 yr ago (2020 August 26). This work was supported by National SKA Program of China (2022SKA0110202) and China Manned Space Program through its Space Application System.
\end{acknowledgements}
\newpage
\bibliography{RAA-2024-0362}
\bibliographystyle{raa}

\end{document}